\documentclass[pre,twocolumn]{revtex4}
\usepackage{epsfig}
\usepackage{bm}
\usepackage{amsmath}
\usepackage{hyperref}
\usepackage{breakurl}

\begin{document}

\title{Chaotic and turbulent temperature fluctuations in atmospheric free convection}

\author{A. Bershadskii}

\affiliation{
ICAR, P.O. Box 31155, Jerusalem 91000, Israel
}

\begin{abstract}

 It is shown, using results of direct numerical simulations, laboratory experiments, measurements in the atmospheric boundary layer, satellite infrared radiances data, the planetary-scale surface-air and tropospheric data, that the temperature fluctuations in atmospheric free convection can be well described by the spatio-temporal distributed chaos approach based on the Bolgiano-Obukhov phenomenology.
 A possibility of analogous consideration of the atmospheric dynamics on other planets (Mars, Jupiter and Saturn) has been briefly discussed using the data obtained with the Phoenix lander and the infrared observations.
\end{abstract}

\maketitle

\section{Inroduction}

At free convection the fluid or gas are driven by the density fluctuations produced by temperature (humidity) differences (see, for instance, Ref. \cite{fok}). The mean flows in forced convection can drastically change the fluid dynamics and result in non-universality of the observed spectra of the atmospheric temperature fluctuations. On the other hand, one can expect that the free convection can be characterized by universal types of the spectra. Therefore, the atmospheric free convection can be used for the fundamental studies of the buoyancy driven turbulence. Moreover, the near free convection can play a significant role in the large-scale atmospheric motions (on the planetary scales), where the winds can be considered as a natural part of the buoyancy driven atmospheric free convection itself.  \\

  The deterministic chaos in the atmospheric motions was also discovered on a simple model of free convection \cite{lorenz} (see for a recent review Ref. \cite{van}). In present paper the deterministic chaos approach has been generalized on the case with randomly fluctuating parameters and takes the form of distributed chaos, that allows consideration of thermal convection with large Rayleigh (Reynolds) numbers typical for atmosphere. This consideration is based on the Bolgiano-Obukhov phenomenology \cite{bol}-\cite{ch} and supported by comparison with direct numerical simulations, laboratory experiments and atmospheric measurements in a wide range of the spatio-temporal scales. \\
  
  In Section II the free convection in the Boussinesq approximation and the Kolmogorov-Bolgiano-Obukhov approach for the inertial-buoyancy range have been considered both for unstable and stable stratification. In Section III the Bolgiano-Obukhov spatial distributed chaos has been introduced. In Section IV results of laboratory experiments for large Rayleigh numbers have been discussed and compared with the predictions made for the Bolgiano-Obukhov spatial distributed chaos. In Section V these predictions have been compared with results of measurements made in the atmospheric boundary layer (over sea and land) and with results of satellite infrared radiances measurements over a planetary-scale tropic area. In Section VI the Bolgiano-Obukhov temporal distributed chaos has been introduced. In Section VII the planetary-scale data for the surface-air (over land) and tropospheric temperature fluctuations (for Arctic) have been discussed and compared with the predictions made for the Bolgiano-Obukhov temporal distributed chaos. In Sections VIII-X an analogous consideration for the Mars's, Jupiter's and Saturn's atmospheres has been briefly discussed using the data obtained with the Phoenix lander, the recent MACDA Mars reanalysis of the data inferred from the Thermal Emission Spectrometer on board of the Mars Global Surveyor satellite, with the images acquired by the Jovian Infrared Auroral Mapper onboard the Juno mission and the data obtained with the NASA Infrared Telescope on Mauna Kea.

\section{Free thermal convection}

  The buoyancy driven convection in the Boussinesq approximation is described by equations \cite{kcv}
$$
\frac{\partial {\bf u}}{\partial t} + ({\bf u} \cdot \nabla) {\bf u}  =  -\frac{\nabla p}{\rho_0} + \sigma g \theta {\bf e}_z + \nu \nabla^2 {\bf u}   \eqno{(1)}
$$
$$
\frac{\partial \theta}{\partial t} + ({\bf u} \cdot \nabla) \theta  =  S  \frac{\Delta}{H}e_z u_z + \kappa \nabla^2 \theta, \eqno{(2)}
$$
$$
\nabla \cdot \bf u =  0 \eqno{(3)}
$$
where $\theta$ is the temperature fluctuations over a temperature profile $T_0 (z)$, $p$ is the pressure, ${\bf u}$ is the velocity, ${\bf e}_z$ is a unit vector along the gravity direction, $g$ is the gravity acceleration, $H$ is the distance between the layers and $\Delta$ is the temperature difference between the layers, $\rho_0$ is the mean density, $\nu$ is the viscosity and $\kappa$ is the thermal diffusivity, $\sigma$ is the thermal expansion coefficient.  For the unstable stratification $S=+1$ and for the stable stratification $S=-1$.\\   

  These equations at $\nu=\kappa=0$ have an invariant 
$$
\mathcal{E} = \int_V ({\bf u}^2 -S\sigma g \frac{H}{\Delta}\theta^2) ~ d{\bf r}   \eqno{(4)}
$$    
where $V$ is the spatial volume (see, for instance, Ref. \cite{kcv}). It can be readily shown that presence of a global rotation does not change the invariant Eq. (4).\\
  
  Corresponding generalization of the Kolmogorov-Bolgiano-Obukhov approach for the inertial-buoyancy range of the spatial scales \cite{my} provides a relationship between characteristic temperature fluctuations $\theta_c$ and the characteristic wavenumber scale $k_c$
$$
\theta_c\propto  (\sigma g)^{-1} \varepsilon^{2/3} k_c^{1/3} \eqno{(5)}
$$
where the generalized dissipation/transfer rate 
$$
\varepsilon = \left|\frac{d\langle{\bf u}^2 -S\sigma g \frac{d}{\Delta}\theta^2 \rangle}{dt}\right| \eqno{(6)}
$$
$\langle ... \rangle$ denotes spatial averaging.\\

  When the buoyancy forces dominate over the inertial forces (see for the stable stratification Ref. \cite{my} and for the unstable one Refs. \cite{pz}-\cite{fl}) one can use 
$$
\varepsilon_b = \left|\frac{d\langle \theta^2 \rangle}{dt}\right| \eqno{(7)}
$$  
instead of the $\varepsilon$ and then one obtains
$$
\theta_c \propto (\sigma g)^{-1/5} \varepsilon_b^{2/5} k_c^{1/5}  \eqno{(8)}
$$
instead of the relationship Eq. (5).\\
\begin{figure} \vspace{-1.8cm}\centering
\epsfig{width=.43\textwidth,file=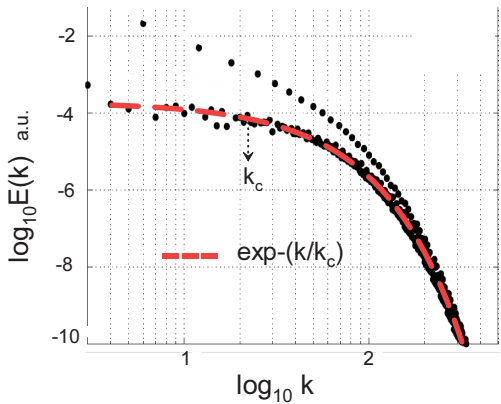} \vspace{-3.05cm}
\caption{Power spectrum of the temperature fluctuations at $Pr =1$ and $Ra = 6.6 \times 10^6$ (onset of the convective turbulence).} 
\end{figure}
\begin{figure} \vspace{-1cm}\centering
\epsfig{width=.45\textwidth,file=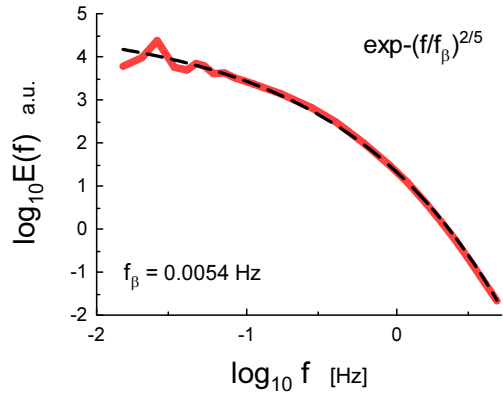} \vspace{-4.5cm}
\caption{Power spectrum of the experimentally observed temperature fluctuations at $Pr = 0.8$ and $Ra = 1.35 \times 10^{14}$. } 
\end{figure}
\begin{figure} \vspace{-0.5cm}\centering \hspace{-1cm}
\epsfig{width=.465\textwidth,file=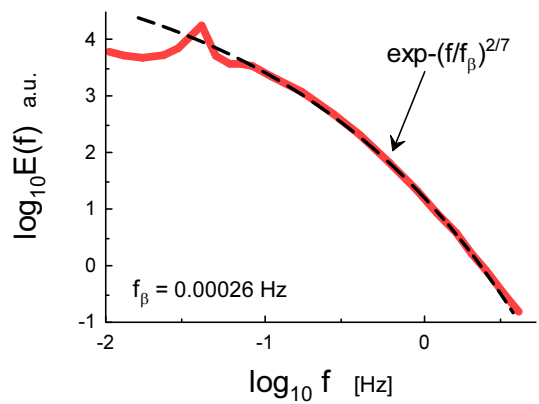} \vspace{-4cm}
\caption{The same as in Fig. 2 but for $Ra = 2.05 \times 10^{11}$.} 
\end{figure}

\section{Spatial distributed chaos} 

  For the bounded and smooth dynamical systems the exponential frequency spectrum 
$$ 
E(f) \propto \exp(-f/f_c)  \eqno{(9)}
$$
 is usually associated with deterministic chaos (see, for instance, Refs. \cite{mm}-\cite{fm} and Section VI).
 
  For the spatial (wavenumber) domain this spectrum is replaced by the spectrum 
$$ 
E(k) \propto \exp(-k/k_c)  \eqno{(10)}
$$ 
(see, for instance, Ref. \cite{mm} and references therein).\\

  In the buoyancy driven unstably stratified thermal convection at the Prandtl number $Pr =\nu/\kappa \sim 1$ transition to turbulence occurs at the Rayleigh number $Ra \sim 10^6$ \cite{v}. Figure 1 shows (in the log-log scales) power spectrum of temperature fluctuations at $Pr =1$ and $Ra = 6.6 \times 10^6$ obtained in a direct numerical simulation reported in Ref. \cite{mv} (the spectral data were taken from the Fig. 11 of the Ref. \cite{mv}). Two branches of this spectrum correspond to different number of the Fourier modes: the upper branch to a small number whereas the lower branch corresponds to the most of the modes. Therefore, the only lower one has a physical significance. The dashed curve in the Fig. 1 indicates the exponential spectrum Eq. (10). The dotted arrow indicates position of the $k_c$.\\

  Transition to turbulence usually results in fluctuations of the parameter $k_c$. These fluctuations can be taken into account by an ensemble averaging
$$
E(k) \propto \int_0^{\infty} P(k_c) \exp -(k/k_c)dk_c \propto \exp-(k/k_{\beta})^{\beta} \eqno{(11)}
$$  
The stretched exponential in the right-hand side of the Eq. (11) is generalization of the exponential spectrum Eq. (10). An estimation of the asymptotic (at large $k_c$) behaviour of the probability distribution $P(k_c)$ 
$$
P(k_c) \propto k_c^{-1 + \beta/[2(1-\beta)]}~\exp(-\gamma k_c^{\beta/(1-\beta)}) \eqno{(12)}
$$     
(where $\gamma$ is a constant) can be made from the Eq. (11) \cite{jon}.\\

  The Eqs. (5) and (8) can be considered in a general form
$$
\theta_c \propto  k_c^{\alpha}   \eqno{(13)}
$$

 When $\theta_c$ has a Gaussian distribution  \cite{my} relationship between parameters $\alpha$ and $\beta$ 
$$
\beta = \frac{2\alpha}{1+2\alpha}  \eqno{(14)}
$$
follows immediately from the Eqs. (12-13). \\

    For the Eq. (5) $\alpha =1/3$,  hence  $\beta = 2/5$ and
$$
E(k) \propto \exp-(k/k_{\beta})^{2/5}.  \eqno{(15)}
$$  
  
  For the Eq. (8) $\alpha =1/5$,  hence  $\beta = 2/7$ and
$$
E(k) \propto \exp-(k/k_{\beta})^{2/7},  \eqno{(16)}
$$
\begin{figure} \vspace{-1.35cm}\centering
\epsfig{width=.45\textwidth,file=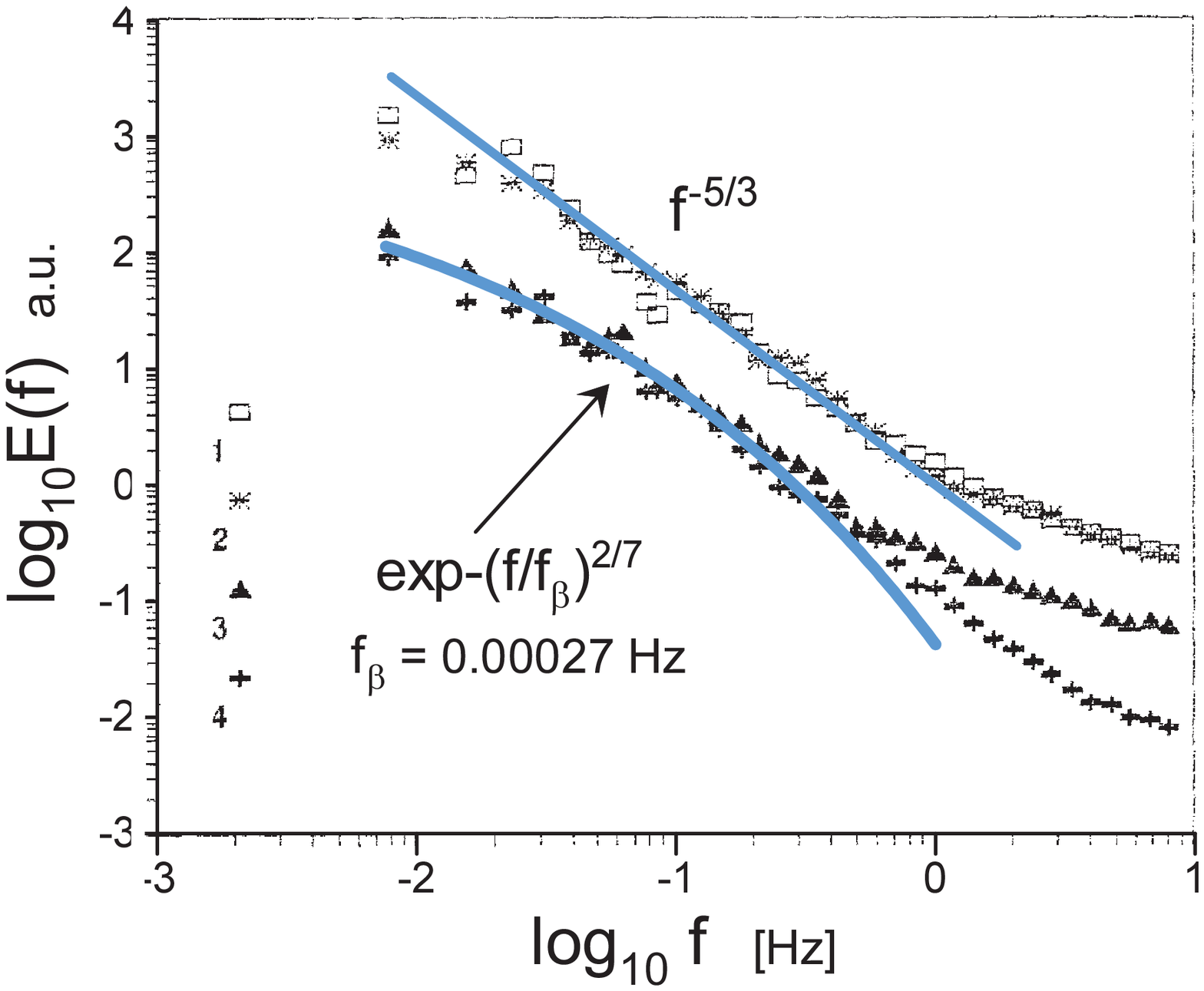} \vspace{-4.1cm}
\caption{Four power spectra of the temperature fluctuations in the atmospheric surface layer over sea at near free convection conditions.} 
\end{figure}
\begin{figure} \vspace{-0.5cm}\centering
\epsfig{width=.45\textwidth,file=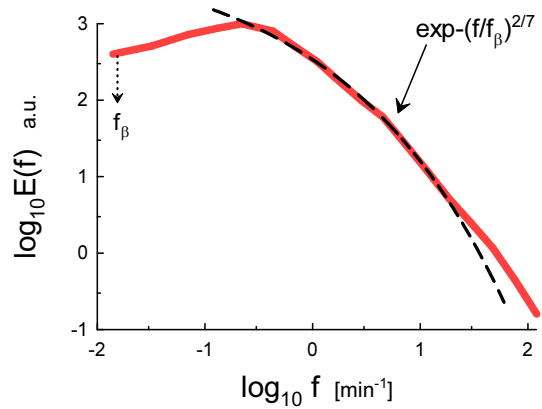} \vspace{-3.85cm}
\caption{Power spectra of the temperature fluctuations in the atmospheric surface layer over land at near free convection conditions.} 
\end{figure}

\section{Laboratory experiments}

   In the laboratory experiments the frequency spectra are usually obtained instead of the wavenumber ones. Figure 2, for instance, shows power spectrum of the experimentally observed temperature fluctuations for the Rayleigh-B\'{e}nard  convection (unstable stratification) in a cylindrical cell with $Pr = 0.8$ and at very high  Rayleigh number $Ra = 1.35 \times 10^{14}$ (the spectral data were taken from Fig. 2 of the Ref. \cite{he1}).\\ 
  
   This is a frequency spectrum. However, in the case of free convection the mean flow velocity $U_0$, usually used for the Taylor hypothesis \cite{my},\cite{mm} relating $f$ and $k$:
$f = U_0~k/2\pi$, can be replaced by a characteristic velocity of advection by the energy-containing eddies past the probe \cite{t},\cite{kv} (see also Ref. \cite{b} and references therein) and we should compare this spectrum with the Eq. (15). The dashed curve in the Fig. 2 indicates correspondence to the stretched exponential Eq. (15).\\ 
  
   Figure 3 shows power spectrum of the temperature fluctuations observed in analogous experiment but for considerably smaller Rayleigh number $Ra = 2.05 \times 10^{11}$. The dashed curve in the Fig. 3 indicates correspondence to the stretched exponential Eq. (16). 

\section{Atmospheric thermal convection}

  The Ref. \cite{gr} reports results of measurements of the temperature fluctuations in the near free convection (at unstable stratification) by a fixed probe over sea surface (the height of the tower was $\sim$ 12m above the sea surface). The weather was calm and the sea surface was aerodynamically smooth. In the air the r.m.s. horizontal velocity fluctuations were comparable to the wind gusts. Results for four data sets obtained during a day were reported: two with comparatively high temperature fluctuations and two with low ones. Figure 4 shows the four power spectra of the temperature fluctuations corresponding to the four data sets (the spectral data were taken from Fig. 1c of the Ref. \cite{gr}). The solid straight line is drawn to indicate the "-5/3" power law for the strong fluctuations case \cite{my}, whereas the curved solid line is drawn to indicate the stretched exponential Eq. (16) for the weak fluctuations case (see previous Section about applicability of the Taylor hypothesis for the free convection). \\

  The Ref. \cite{emf} reports results of measurements of the temperature fluctuations in the near free convection over land. The measurements were produced by 'Eddy-covariance' tower (at 2.29m height) and Sodar/RASS system in the morning hours (from 07:35 to 08:40 UTC) over a corn field.  Moderate to high buoyancy fluxes and a drop of the wind speed result in domination of buoyancy over shear. The occurrence of large-scale turbulence structures (plumes) were registered. The free convection conditions occur at about 50 percent of the 92 days of the measurement period. Figure 5 shows power spectrum of the temperature fluctuations (the sonic spectral data were taken from Fig. 6h of the Ref. \cite{emf}). The dashed line is drawn to indicate the stretched exponential Eq. (16) and the dotted arrow indicates position of $f_{\beta}$. \\
  
\begin{figure} \vspace{-1.3cm}\centering
\epsfig{width=.45\textwidth,file=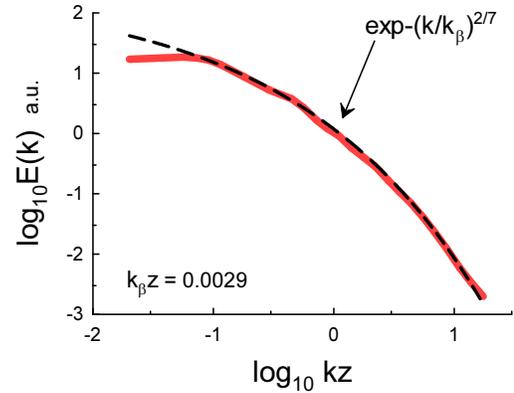} \vspace{-4.5cm}
\caption{Power spectra of the alongwind-sampled temperature fluctuations in the unstable atmospheric surface layer over ocean.} 
\end{figure}  
\begin{figure} \vspace{-0.5cm}\centering
\epsfig{width=.43\textwidth,file=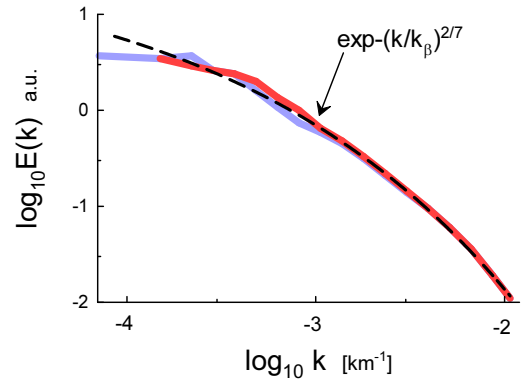} \vspace{-4cm}
\caption{Power spectra of the MTSAT satellite thermal infrared radiances (the two different colors correspond to zonal and  meridional spectra).} 
\end{figure}

    It is well known (although not well understood) that the free-like convection behaviour can be often observed under conditions which seem to be rather different from the free ones. Figure 6, for instance, shows power spectrum of the temperature fluctuations airborne measured in the Weastern Atlantic ocean at 50m level above the water surface during cold air outbreaks \cite{cy}. The measurements were made in near-shore areas of roll vortices and cloud streets under considerable mean wind shear. The alongwind-sampled spectral data were taken from Fig. 4 of the Ref. \cite{cy} ($k$ is the alongwind wavenumber, $z$ is the height above the ocean surface). The wavenumber was calculated using the Taylor hypothesis $k = 2\pi f/V_a$, where $V_a$ is the relative velocity between the aircraft and air. The dashed line is drawn to indicate the stretched exponential Eq. (16). \\

    The geostationary satellite infrared radiances data for spatial range of scales from 100km up to 5000km can also be viewed as a result of a nearly free convection. At this case the winds can be approximately considered as an internal part of the large-scale thermal convection. \\
    
    In the case of the planetary scales one should replace the plane horizontal layer by a thin spherical shell and take into account the global rotation term in the Eq. (1) (both these amendments conserve the generalised energy Eq. (4)). The global radiation and water in different phases (clouds etc.) are much more difficult to account. However, all these additional factors can be integrated into a generalized dissipation/transfer rate and the Kolmogorov-Bolgiano-Obukhov like estimates Eq. (5) and (8) can be still preserved for an inertial-buoyancy range of scales.\\

   Figure 7 shows wavenumber power spectrum corresponding to these conditions (the spectral data were taken from Fig. 2 of Ref. \cite{pls}). The spectra were computed using 1386 images from the infrared channel sensitive to temperature at the top of clouds over area with longitudes 80$^o$E-200$^o$E and latitudes 40$^o$S-30$^o$N and with resolution 30 km. The data were obtained by the geostationary MTSAT satellite (see Ref. \cite{tak} and references therein). The dashed line in the Fig. 7 is drawn to indicate the stretched exponential Eq. (16).\\
   
\begin{figure} \vspace{-1.6cm}\centering
\epsfig{width=.45\textwidth,file=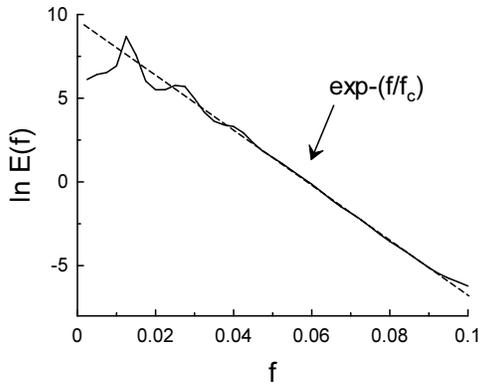} \vspace{-4.15cm}
\caption{Power spectrum of $z$-component for the Lorenz system.} 
\end{figure}
\begin{figure} \vspace{-1.1cm}\centering
\epsfig{width=.48\textwidth,file=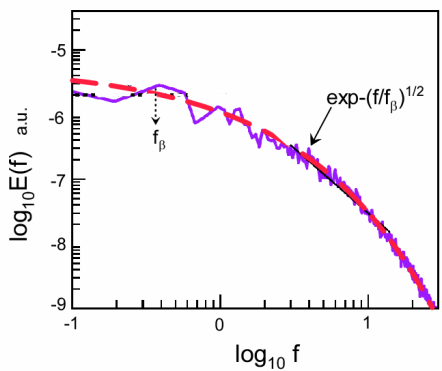} \vspace{-4.9cm}
\caption{Power spectrum (in the log-log scales) of the pure temporal temperature fluctuations for a DNS of the Rayleigh-B\'{e}nard convection. } 
\end{figure}
\begin{figure} \vspace{-1.45cm}\centering
\epsfig{width=.45\textwidth,file=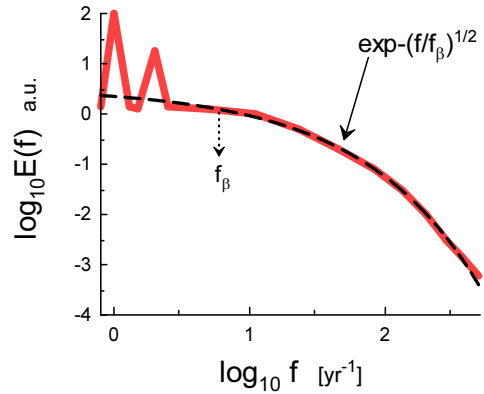} \vspace{-4.3cm}
\caption{Power spectrum of temperature fluctuations obtained with the Twentieth Century Reanalyses (20CR \cite{com}) for Arctic troposphere.} 
\end{figure}
    
\section{Temporal distributed chaos}

  It was already mentioned in the Section III that  for the bounded and smooth dynamical systems the exponential frequency spectrum Eq. (9) is usually associated with deterministic chaos. A very simplified (on the first Galerkin modes) model of the Rayleigh-B\'{e}nard convection (the Lorenz system \cite{lorenz})
$$
\frac{dx}{dt} = \sigma (y - x),~~      
\frac{dy}{dt} = r x - y - x z, ~~
\frac{dz}{dt} = x y - b z,             \eqno({17})   
$$
exhibits such spectrum for the parameters $b = 8/3,~ r = 28.0, ~\sigma=10.0$, for instance. Figure 8 shows power spectrum of $z$-component for this case. The dashed straight line corresponds to the Eq. (9) in the semi-logarithmic scales. \\  

For more complex (real) situations the parameter $f_c$ can strongly fluctuate and it becomes a necessity to consider an ensemble average over this fluctuating parameter in order to compute the power spectrum 
$$
E(f) = \int P(f_c) ~\exp-(f/f_c)~ df_c, \eqno{(18)}
$$  

  In the true temporal case the estimate Eq. (5) should be replaced by relationship
$$
\theta_c\propto  (\sigma g)^{-1} \varepsilon^{1/2} f_c^{1/2}, \eqno{(19)}
$$

\begin{figure} \vspace{-1.8cm}\centering
\epsfig{width=.45\textwidth,file=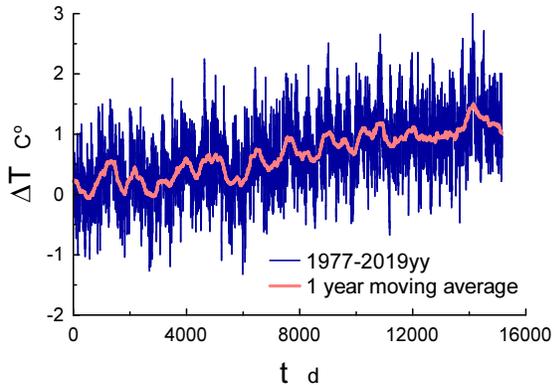} \vspace{-4.3cm}
\caption{Global average surface temperature anomaly for 1977-2019yy. The land-based daily data were taken from the Ref. \cite{ber2}.} 
\end{figure}
\begin{figure} \vspace{-0.1cm}\centering
\epsfig{width=.47\textwidth,file=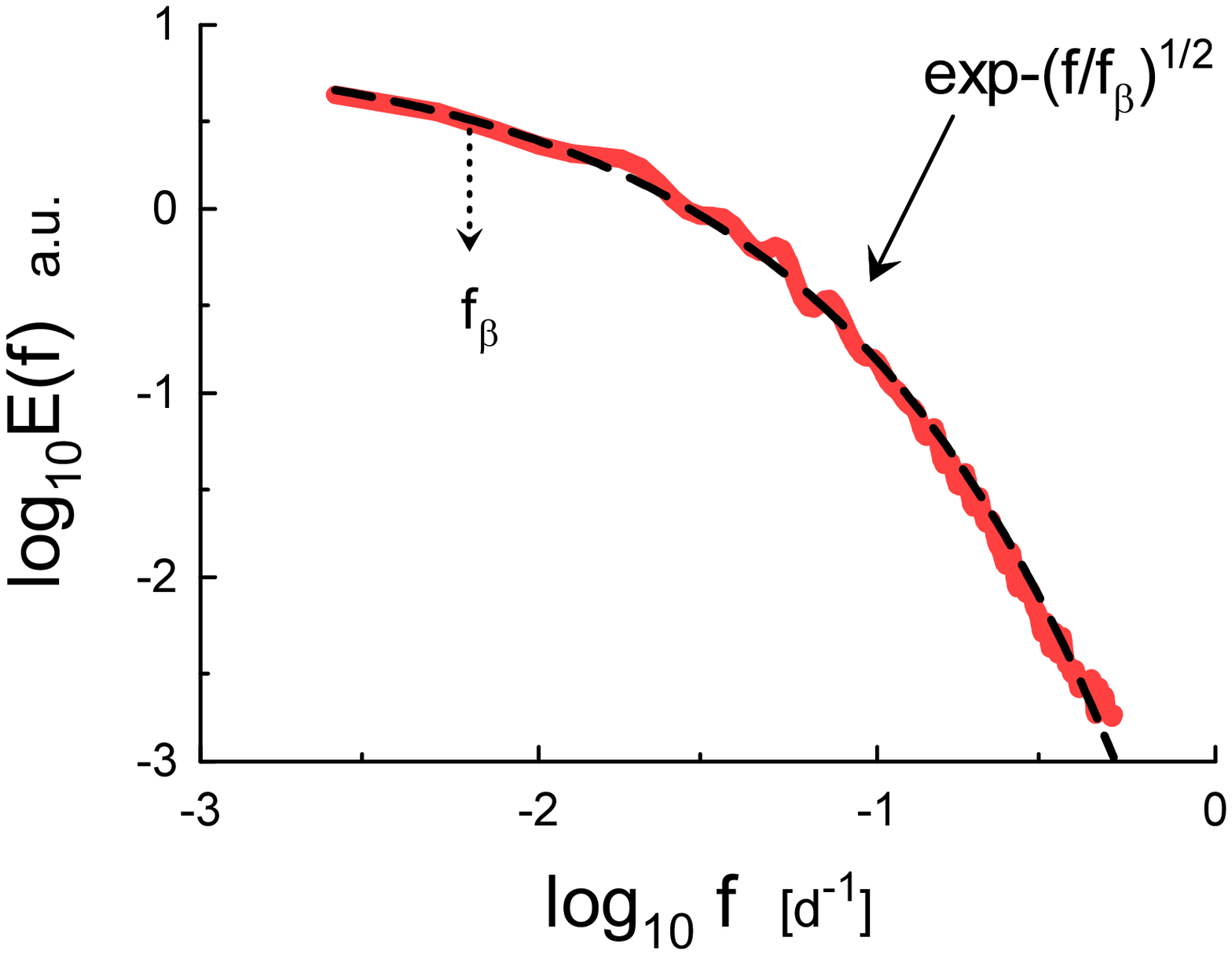} \vspace{-3.9cm}
\caption{Power spectrum corresponding to the detrended daily data for the period 1977-2019yy.} 
\end{figure}
\begin{figure} \vspace{-1.42cm}\centering
\epsfig{width=.45\textwidth,file=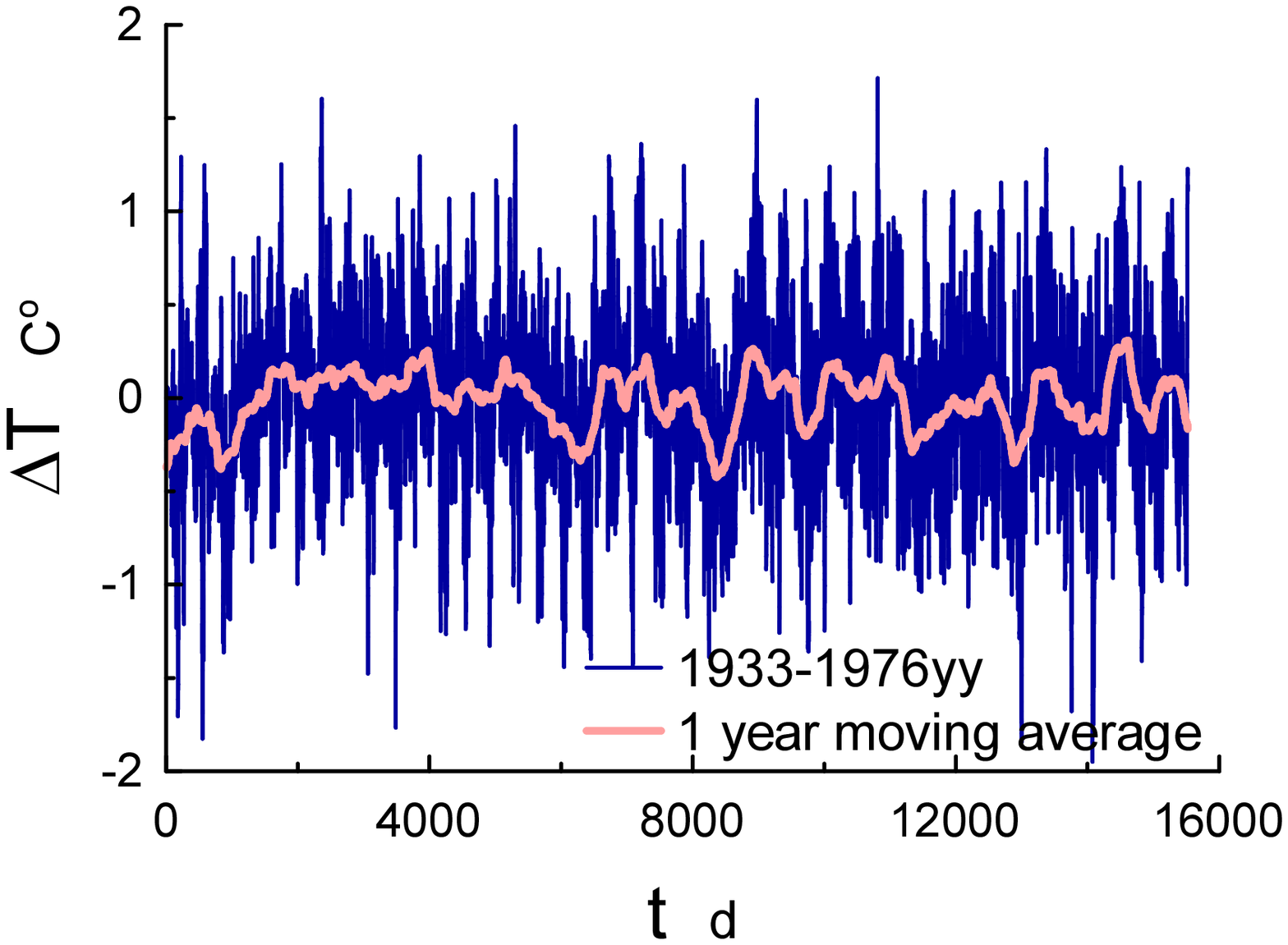} \vspace{-4.7cm}
\caption{Global average surface temperature anomaly for 1933-1976yy. The land-based daily data were taken from the Ref. \cite{ber2}.} 
\end{figure}
\begin{figure} \vspace{+0.43cm}\centering
\epsfig{width=.47\textwidth,file=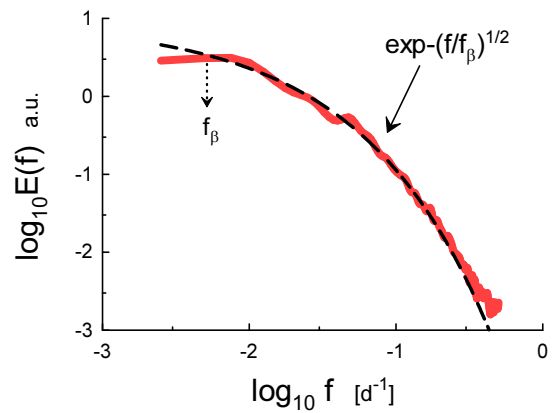} \vspace{-4.45cm}
\caption{Power spectrum corresponding to the detrended daily data for the period 1933-1976yy.} 
\end{figure}

  Then, the distribution $P(f_c)$ can be readily calculated in the frames of the Kolmogorov-Bolgiano-Obukhov phenomenology from the Eq. (19) if the characteristic temperature fluctuations $\theta_c$ is normally distributed
$$
P(f_c) \propto f_c^{-1/2} \exp-(f_c/4f_{\beta})  \eqno{(20)}
$$
where $f_{\beta}$ is a renormalised, due to the fluctuations, characteristic frequency (cf. the Eq. (21)).

     Substituting the Eq. (20) into the Eq. (18) we obtain
$$
E(f) \propto \exp-(f/f_{\beta})^{1/2}  \eqno{(21)}
$$

 Figure 9 shows a pure temporal (frequency) power spectrum for the temperature fluctuations obtained in direct numerical simulations (DNS) of the system Eqs. (1-3) (the Rayleigh-B\'{e}nard convection). The spectral data were taken from Fig. 7 of the the Ref. \cite{kv}. The measurements in these DNS were made with the real-space probes located at the centre of a box. The mean velocity at the centre of the cubical box has zero value. Therefore, the measurements provide real temporal (frequency) spectrum \cite{kv}. The Prandtl number was $Pr = 1$ (i.e. close to the usual for the air at normal conditions) and the Rayleigh number was taken $Ra = 10^8$. \\

  The dashed curve in the Fig. 9 indicates the stretched exponential spectrum Eq. (21) and the dotted vertical arrow indicates location of  the frequency $f_{\beta}$.  \\
   
\section{Surface-air and tropospheric temperature fluctuations on the planetary scales} 

  For the planetary scales the winds can be considered as a part of the planetary-scale thermal convection. Therefore one can expect that the free thermal convection consideration can be applied for these scales. \\
\begin{figure} \vspace{-1.6cm}\centering
\epsfig{width=.45\textwidth,file=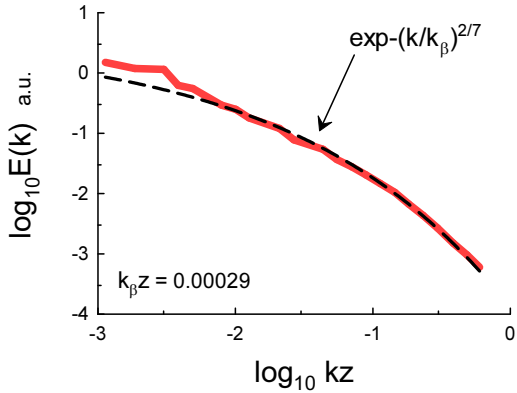} \vspace{-4.25cm}
\caption{Power spectrum of the alongwind temperature fluctuations observed in the Martian atmospheric boundary layer.} 
\end{figure}
\begin{figure} \vspace{-0.5cm}\centering
\epsfig{width=.45\textwidth,file=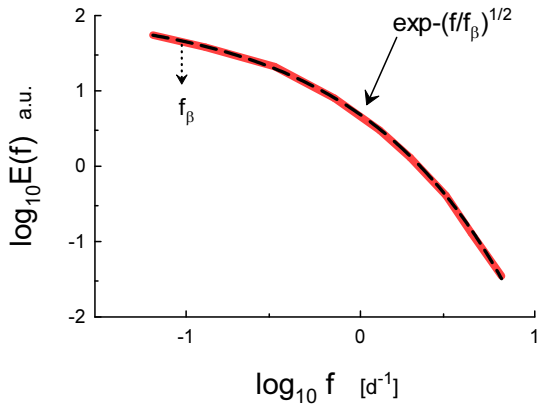} \vspace{-4.25cm}
\caption{Power spectrum of the global surface-atmospheric temperature fluctuations for Mars.} 
\end{figure}
  In the case of the planetary-scale thermal convection one should replace the plane horizontal layer by a thin spherical shell and take into account the global rotation term in the Eq. (1) (both these amendments conserve the generalised energy Eq. (4)). The global radiation and water in different phases (clouds etc.) are much more difficult to account. However, all these additional factors can be integrated into a generalized dissipation/transfer rate $\varepsilon $ and the Kolmogorov-Bolgiano-Obukhov like estimate Eq. (19) can be still preserved for an inertial-buoyancy range of scales.\\
  
  Figure 10 shows power spectrum of temperature fluctuations obtained with the Twentieth Century Reanalyses - 20CR \cite{com} at altitude 700mb (approximately 3013m). The spectra at the latitude $75^o$ N (Arctic) have been averaged over all $180^o$ longitude ($2^o\times 2^o$ elements). The spectral data have been taken from Fig. 1a of the Ref. \cite{ls}. The two peaks correspond to the the annual frequency and its first harmonic. The dashed curve in the Fig. 10 indicates the stretched exponential spectrum Eq. (21) and the dotted vertical arrow indicates location of  the frequency $f_{\beta}$.\\

  The good quality daily historical time series of the global surface-air temperature over land are now available for a statistical analysis \cite{ber1},\cite{ber2}. 
 
   Figure 11 shows a global surface-air temperature anomaly for a recent time period 1977-2019yy (relative to the Jan. 1951 - Dec. 1980 average). The daily data (land-based) for this period were taken from the site Ref. \cite{ber2}. Since the data are statistically non-stationary a certain detrending is necessary before a proper spectral analysis. The detrending was produced by subtraction of the one year moving average of the data set (shown as the red line in the Fig. 11) from the original one. This subtraction removes the long-term trends and the remaining time-series represent the daily
to intra-annual dynamics only (cf. Refs. \cite{wh},\cite{ven2} for such type of detrending).\\

 Figure 12 shows power spectrum corresponding to the detrended data. The power spectrum was calculated
with the maximum entropy method, this method gives an optimal resolution for the short data-sets \cite{oh}. The dashed line in the Fig. 12 indicates the stretched exponential decay Eq. (21). The dotted arrow indicates position of the scale $f_{\beta}$.\\
   
     The data for a previous time period of the same length 1933-1976yy (taken from the same site \cite{ber2}) were detrended by the same method - Fig. 13. Figure 14 shows corresponding power spectrum. The dashed line in the Fig. 14 indicates the stretched exponential decay Eq. (21). The dotted arrow indicates position of the scale $f_{\beta}$.\\
     
\section{Martian atmospheric thermal convection}    

    Main physical properties of the 'climate and meteorology' related to the Martian atmosphere turned out to be surprisingly similar to those of the Earth atmosphere (see for a review Refs. \cite{rlm},\cite{clm}) and references therein). Although the Martian atmosphere (consisting mostly of $\mathrm{CO}_2$) is thinner than that of Earth the thermal (buoyancy driven) convection processes in this atmosphere are no less vigorous and substantial than those in the Earth atmosphere. The main difference in this respect is the strong influence of water (in different phases) on the Earth's atmospheric dynamics, whereas the dust is the rather significant factor in the Martian atmospheric dynamics. Therefore it is interesting to compare the above results obtained for Earth with those available now for Mars . \\
   
   We have already seen in the Earth atmospheric boundary layer (Fig. 6) that  the free-like convection behaviour can be observed under conditions which seem to be rather different from the free ones. Figure 15 shows power spectrum of the alongwind temperature fluctuations observed in the Martian atmospheric boundary layer by the Phoenix lander (the spectral data were taken from Fig. 8 of the Ref. \cite{davy}). The measurements were made at about 2 m height above the Mars surface ($k$ is the alongwind wavenumber, $z$ is the height above the surface, see the Ref \cite{davy} for more details). As for the Fig. 6 the wavenumber was calculated using the Taylor hypothesis $k = 2\pi f/V$, where $V$ is the mean wind speed. The dashed line is drawn to indicate the stretched exponential Eq. (16).\\
   
   Let us now consider the Martian global temperature fluctuations. Sun-synchronous 2-hour polar orbits (there was a $30^o$ displacing in longitude at every new orbit) of the NASA's Mars Global Surveyor with the Thermal Emission Spectrometer on board provided a well-sampled temporal and spatial coverage for the period from February 1999 to August 2004. The MACDA (Mars Analysis Correction Data Assimilation) reanalysis of the data inferred from the Thermal Emission Spectrometer measurements were then used to estimate the Martian surface-atmospheric temperature fluctuations \cite{clm},\cite{MACDA}. 
   
   Figure 16 shows corresponding power spectrum (the spectral data were taken from Fig. 1b of the Ref. \cite{lah}). The strong spectral spikes corresponding to the Mars diurnal cycle and its harmonics are not shown in the Fig. 16 (a day on Mars - 1 sol, is approximately equal to 24.6 hours.). The dashed line in the Fig. 16 indicates the stretched exponential decay Eq. (21). The dotted arrow indicates position of the scale $f_{\beta}$.
   
\begin{figure} \vspace{-1.3cm}\centering
\epsfig{width=.45\textwidth,file=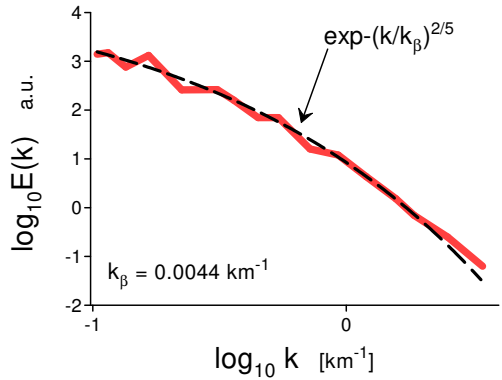} \vspace{-3.78cm}
\caption{Power spectrum of the stratospheric temperature fluctuations for Saturn ($k$ is the vertical wavenumber).} 
\end{figure}
\begin{figure} \vspace{-1.24cm}\centering
\epsfig{width=.45\textwidth,file=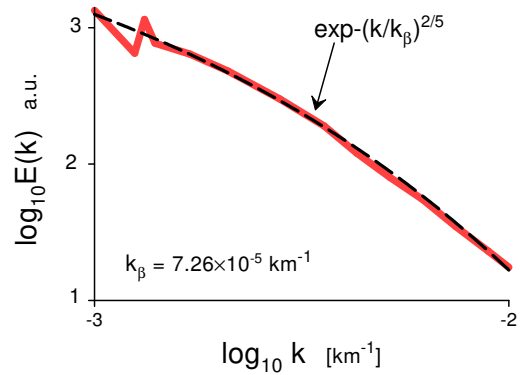} \vspace{-3.84cm}
\caption{Power spectrum for the mid-infrared thermal emission for Jupiter's south polar atmosphere ($k$ is the horizontal wavenumber).} 
\end{figure}
\begin{figure} \vspace{-0.6cm}\centering
\epsfig{width=.49\textwidth,file=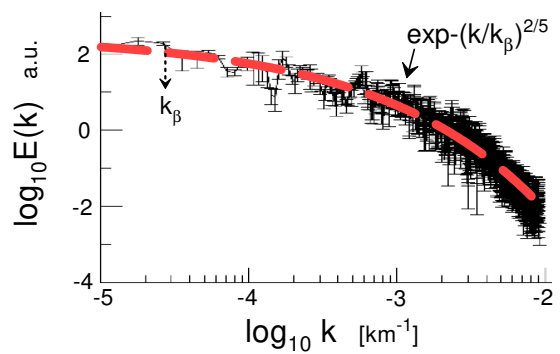} \vspace{-6cm}
\caption{As in the Fig. 18 but for a narrow circular path outside of the cyclones' belt near Jupiter’s South Pole.} 
\end{figure}

\section{Jupiter's and Saturn's atmosphere}

   Although the Jupiter's and Saturn's atmospheres are rather different from that on Earth it seems that the thermal convection also plays very significant role in their atmospheric dynamics.\\
   
  In the Ref. \cite{hfm} a power spectrum of the temperature fluctuations in Saturn's stratosphere was reported. The temperature fluctuations were obtained for the vertical temperature profile at 55.5$^{o}$ S latitude using the NASA Infrared Telescope on Mauna Kea. The inverted occultation method was used for this purpose. Figure 17 shows the power spectrum (the spectral data were taken from Fig. 7b of the Ref. \cite{hfm}). The dashed line in the Fig. 17 indicates the stretched exponential decay Eq. (15). \\

   The recent papers Refs. \cite{adr},\cite{mori} reported an analysis of the data obtained with the Jovian Infrared Auroral Mapper (JIRAM) onboard the Juno mission for the Jupiter’s South Pole atmosphere. Figure 18 shows an example of the power spectra for the mid-infrared (thermal) emission for Jupiter's south polar atmosphere (the latitudes higher than 82$^o$ S). This region was dominated by the polar cyclones. The cyclones are characterized by thick clouds. At the presence of a thick cloud cover the JIRAM is sensible to the temperatures of the cloud top. The spectral data for the Fig. 18 were taken from the Fig. 11c (the upper curve) of the Ref. \cite{adr}. The dashed line in the Fig. 18 indicates the stretched exponential decay Eq. (15).\\

   Figure 19 shows power spectrum for the mid-infrared (thermal) emission obtained with the JIRAM for a narrow circular path outside of the cyclones' belt near Jupiter’s South Pole. For this region the cloud cover is thinner than that for the region dominated by the polar cyclones and the winds have low speed. The spectral data were taken from Fig. 2 of the Ref. \cite{mori}. The dashed line in the Fig. 19 indicates the stretched exponential decay Eq. (15).\\

  The spatial (wavenumber) spectrum Eq. (15) shown in the Figs. 17-19 has its temporal (frequency) equivalent in the form of the spectrum Eq. (21) (cf. Figs. 12, 14 and 16) - the both spectra (spatial and temporal) are dominated by the generalized dissipation/transfer rate $\varepsilon$ Eq. (6) (cf. Eq. (5) and Eq. (19)).

\section{Conclusions}

    The Bolgiano-Obukhov phenomenology appears to be adequate for the atmospheric free convection in the atmospheric boundary layer (both over land and sea) when it is considered in the frames of the distributed chaos approach. Moreover, it can be also applied to the atmospheric data for the temperature fluctuations even in some cases when the free convection conditions are not satisfied and for the planetary scales when the winds can be considered as a part of the planetary-scale thermal convection. The Martian atmospheric boundary layer and global atmospheric thermal convection can be analyzed in the same way, as well as the observed temperature fluctuations in the Jupiter's and Saturn's atmospheres.

\section{Acknowledgement}

I thank K.R. Sreenivasan for stimulating discussion, and I gratefully acknowledge use of the data provided by the Berkeley Earth dataset.

\end{document}